\documentclass[5p,times]{myelsarticle}

\usepackage[bookmarks,raiselinks,pageanchor,colorlinks,linkcolor=black,citecolor=black,urlcolor=magenta,filecolor=cyan,menucolor=red,pagebackref]{hyperref} 
\usepackage{ecrc}
\volume{00}
\firstpage{1}
\journalname{Journal of Power Sources}
\runauth{J. Stamm et al.}

\jnltitlelogo{J. Power Sources}

\usepackage{amssymb}
\usepackage{amsthm}
\usepackage{color}

\usepackage[figuresright]{rotating}

\usepackage{amsmath}
\usepackage{bbm}
\usepackage{romannum}
\usepackage[intoc]{nomencl}
\usepackage{hyperref}
\makenomenclature
\usepackage{tabu}
\newcounter{chemequation}
\renewcommand*\thechemequation{\Roman{chemequation}}
\newcommand*\chemtag{%
\stepcounter{chemequation}%
\tag{\thechemequation}}

\begin{document}
\pagenumbering{arabic}
\begin{frontmatter}

\dochead{}
\title{Modeling Nucleation and Growth of Zinc Oxide During Discharge of Primary Zinc-Air Batteries}
\author[HIU,DLR,Mun]{Johannes Stamm}
\author[HIU,KIT]{Alberto Varzi}
\author[HIU,DLR,Ulm]{Arnulf Latz}
\author[HIU,DLR]{Birger Horstmann\corref{cor1}}
\ead{birger.horstmann@dlr.de}
\cortext[cor1]{Corresponding author}
\address[HIU]{Helmholtz Institute Ulm (HIU), Helmholtzstra\ss e 11, 89081 Ulm, Germany}
\address[DLR]{German Aerospace Center (DLR), Institute of Engineering Thermodynamics, Pfaffenwaldring 38-40, 70569 Stuttgart, Germany}
\address[Mun]{Institute for Computational and Applied Mathematics, Universit\"at M\"unster, Einsteinstra\ss e 62, 48149 M\"unster, Germany}
\address[KIT]{Karlsruhe Institute of Technology (KIT), PO Box 3640, 76021 Karlsruhe, Germany}
\address[Ulm]{Ulm University, Institute of Electrochemistry, Albert-Einstein-Allee 47, 89069 Ulm, Germany}
\begin{abstract}
Metal-air batteries are among the most promising next-generation energy storage devices. Relying on abundant materials and offering high energy densities, potential applications lie in the fields of electro-mobility, portable electronics, and stationary grid applications. Now, research on secondary zinc-air batteries is revived, which are commercialized as primary hearing aid batteries. One of the main obstacles for making zinc-air batteries rechargeable is their poor lifetime due to the degradation of alkaline electrolyte in contact with atmospheric carbon dioxide. In this article, we present a continuum theory of a commercial Varta PowerOne button cell. Our model contains dissolution of zinc and nucleation and growth of zinc oxide in the anode, thermodynamically consistent electrolyte transport in porous media, and multi-phase coexistance in the gas diffusion electrode. We perform electrochemical measurements and validate our model. Excellent agreement between theory and experiment is found and novel insights into the role of zinc oxide nucleation and growth and carbon dioxide dissolution for discharge and lifetime is presented. {We demonstrate the implications of our work for the development of rechargeable zinc-air batteries.}
\\
\\
\emph{Highlights}
\begin{itemize}
	\item Modeling and simulating of VARTA button cell
	\item Validation of galvanostatic discharge and lifetime analysis
	\item Nucleation and growth of ZnO and its impact on discharge curve
	\item Degradation due to carbonation of alkaline electrolyte
\end{itemize}
\end{abstract}
\end{frontmatter}

\section{Introduction}
\label{sec:intro}
Energy production by renewable energies, i.e., wind or solar power, is fluctuating. Therefore, special efforts are required to match energy production and consumption. Traditional power plants are not ideal to compensate for energy fluctuations, especially because renewable energies are strongly decentralized. Furthermore, portable electronic devices and electro-mobility rely on compact energy storage devices. Metal-air batteries are promising candidates to fulfill this demand, because of their high specific energy density and the use of cheap and abundant materials. These batteries are open at the cathode and use atmospheric oxygen. 

Several metals, e.g., lithium, sodium, and zinc, are potential active anode materials in metal-air cells \cite{Li2014a}. The high theoretical energy density of lithium-air batteries has stimulated a lot of research \cite{Grande2014}. For aprotic electrolytes, the challenge is to influence growth mechanisms in order to maximize capacity, while maintaining sufficient reversibility
\cite{Aetukuri2014, Johnson2014, Liu2015, Lu2016, Gao2016, Schwenke2015, Horstmann2013a}. Aqueous lithium-air batteries require a stable lithium conducting anode protection \cite{Stevens2010, Zhang2011, Horstmann2013, Danner2014}. Non-aqueous sodium-air cells rely on cheap materials having similar challenges as lithium-air batteries \cite{Adelhelm2015,Hartmann2013,Kim2016}. 

Zinc-air batteries stand out as the single commercialized metal-air battery. Primary zinc-air button-cells have a long history in hearing aids. Therefore, also rechargeable zinc-air batteries are in a very mature state \cite{Li2014a,Xu2015,Harting2012,Li2013,Neburchilov2010,McLarnon1991}. The discharge product is not passivating and crystallization is reversible. Metallic zinc (Zn) anodes are stable in aqueous electrolytes and can withstand a few hundred cycles. The cells can work with ambient air for a few months. The theoretical specific energy density of zinc-air batteries reaches $1100 \text{ Wh kg}^{-1}$ with respect to the mass of Zn \cite{Li2014a}. The button cell studied in this work delivers the practical energy density $300 \text{ Wh kg}^{-1}$ at $100 \text{ Am}^{-2}$, which is still about three times as high as batteries in modern electric vehicles \cite{Groger2015}. 

Besides its energy density, zinc-air cells offer a couple of additional advantages, e.g., comparatively constant discharge voltage, long storage life, no reaction with water, large abundance of Zn, low costs, and high environmental safety \cite{Lee2011}. However, unsolved issues for secondary zinc-air cells remain, particularly with respect to cycle life and lifetime. Major challenges are passivation due to zinc oxide (ZnO) precipitation, shape changes of metallic Zn during cycling, and sluggish kinetics of oxygen reduction \cite{Li2014a}. 

Furthermore, atmospheric carbon dioxide enters the cell and reacts to carbonate in the electrolyte \cite{Drillet2001, Schroder2014}. This process entails an irreversible reduction of hydroxide concentration, zincate solubility, and electrolyte conductivity. Therefore, without special precautions, the lifetime of alkaline zinc-air batteries is limited to a few months, which is especially troublesome for secondary cells. 

Most research on zinc-air batteries is devoted to improving the alkaline system \cite{Xu2015}, based on modeling \cite{Cogswell2015}, in-situ x-ray measurements \cite{Arlt2014}, and designing nano-materials \cite{Li2013}. Novel research makes use of alternative electrolytes, i.e., aqueous neutral electrolytes \cite{ThomasGoh2014} to mitigate carbonate formation or ionic liquids \cite{Liu2013} to enable reversible Zn deposition. 

In order to improve the cycle life of zinc-air batteries, a better understanding of its elementary processes seems necessary. To address this issue, several models on zinc-air cells are discussed in the literature \cite{Sunu1980, Mao1992, Deiss2002, Schroder2014}, based on the general, macroscopic, and one-dimensional model for porous electrodes proposed by Newman et al. \cite{Newman1962, Newman2004}. 

Sunu and Bennion \cite{Sunu1980} develop a one dimensional, time dependent model of the Zn anode of zinc-air batteries, based on concentrated ternary electrolyte theory \cite{Newman2004}. It is found that electrolyte convection in Zn anodes can lead to a redistribution of Zn inside the anode and into the cathode upon cycling. The redistributed Zn blocks electrolyte pores or electrically shortens the cell. Isaacson et al. \cite{Isaacson1990} discuss a similar, but two dimensional model for Zn electrodes.

Mao and White \cite{Mao1992} extend Sunu's model resolving the separator region. It is found that potassium zincate does not precipitate under realistic conditions \cite{Ko1983}. Deiss et al. \cite{Deiss2002} describe a similar model for secondary zinc-air cells based on dilute solution theory, which reaches a fairly good agreement with experimental discharge curves.

Schr\"{o}der and Krewer \cite{Schroder2014} develop a model of secondary cells, including a gas diffusion electrode and the effect of atmospheric carbon dioxide. As this model is zero-dimensional, it cannot resolve the nonuniform reaction distribution in the Zn anode \cite{Arlt2014}. The model demonstrates the reduction in lifetime due to carbonization of the alkaline electrolyte.

In this paper, we develop a one dimensional model for both porous electrodes. Three-phase-coexistence in the gas diffusion electrode \cite{Horstmann2013, Danner2014} and inhomogeneous reaction distributions in the Zn anode are modeled at the same time. The electrolyte transport model is based on rational thermodynamics taking into account diffusion, migration, and convection \cite{Latz2015,Zausch2010}. For the first time in zinc-air batteries, we model the nucleation and growth of ZnO and its impact on Zn dissolution. The kinetics of carbon dioxide absorption is described as first order reaction based on a microscopic model \cite{Cents2003}. Our model is parametrized and validated with the commercial zinc-air coin cell Varta PowerOne PR44 Type p675 used for hearing aids. We can correlate characteristic features in the discharge curves with specific processes inside the battery, e.g., nucleation of ZnO and diffusion of reactants through ZnO. The limited battery lifetime is explained with carbonation of the electrolyte. 

Our paper is structured as follows: First, we give a brief overview of cell design, composition, and the chemical reactions during the discharge process (see Sec. \ref{sec:znair}). Next, we describe our homogeneous, one-dimensional, continuum cell model (see Sec. \ref{sec:difmigmod}) and its parameterization (see Sec. \ref{sec:parameters}). Then, we discuss galvanostatic discharge (see Sec. \ref{sec:discharge}) and lifetime (see Sec. \ref{sec:lifetime}). In each of these two sections, we compare experiments and simulations. The excellent agreement between theory and experiment allows the discussion of internal battery processes based on simulations. Finally, we summarize our findings in Sec. \ref{sec:conclusion}.

\section{Zinc-Air Button Cell}
\label{sec:znair}
\begin{figure}
	\includegraphics[width=\columnwidth]{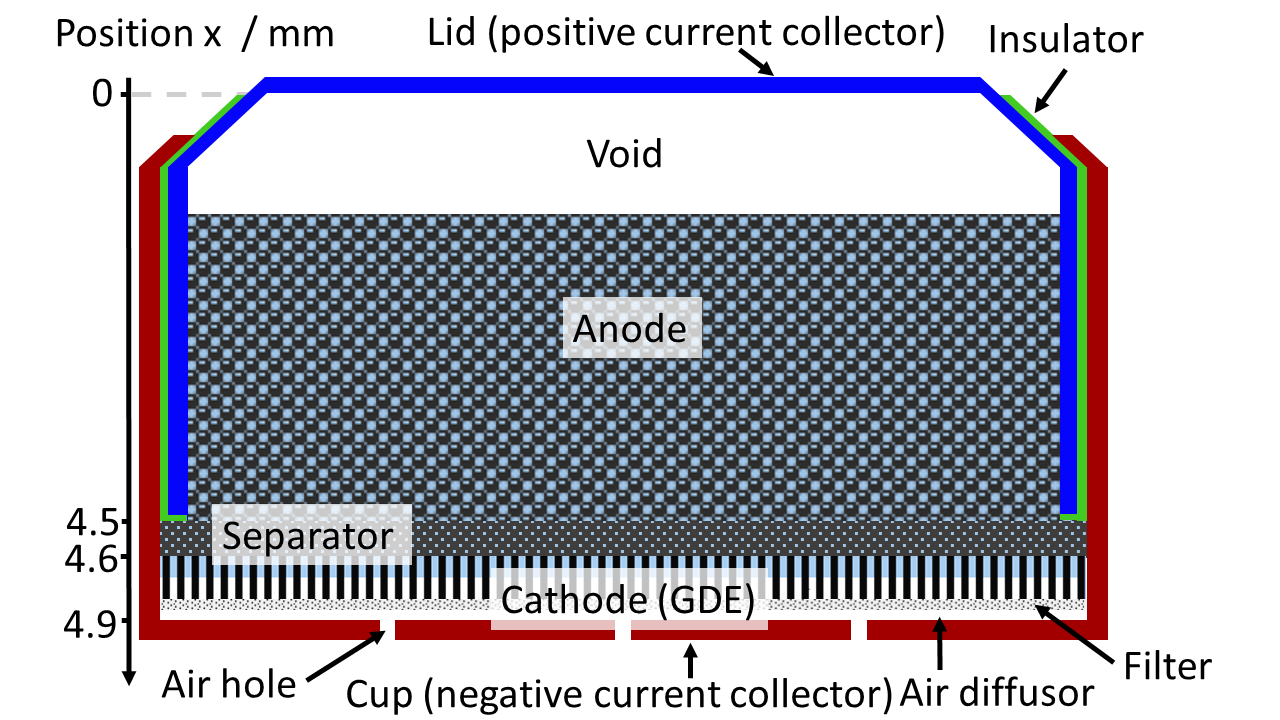}
\caption[Schematic composition of a zinc-air button cell]{Components and structure of a zinc-air button cell.}
\label{fig:aufbau}
\end{figure}
In this section, we describe the structure and components of the Varta PowerOne hearing aid battery PR44 Type p675 (see Fig. \ref{fig:aufbau}). Anode, separator, cathode, and electrolyte are important for cell performance and described in the subsequent sections.  

\subsection{Composition and Design of Zinc-Air Button Cells}
\label{sec:aufbau}
The porous \emph{anode} consists of metallic Zn powder connected to the current collector. The pores are flooded with electrolyte. Zn as active material dissolves during discharge at the solid-liquid phase boundary. The fine powder with its large surface area provides a fast and homogeneous Zn dissolution. A void space beneath the cover of the anode accommodates the volume change due to the conversion of active material in the anode. We assume that this void space is filled with gas at standard pressure which does not interact with the cell, but can leave it through the gas diffusion electrode (GDE).

The \emph{electrolyte} in the VARTA cell is an aqueous potassium hydroxide solution at $32 \text{weight}\%$. This electrolyte is optimized for conductivity realizing the best combination of ionic strength and viscosity. Additionally, this potassium hydroxide solution offers fast oxygen reduction kinetics.

The \emph{separator} is made of a microporous filtering paper. It prevents electric contact between the two electrodes, but allows the electrolyte to pass through.  
 
As \emph{cathode}, metal-air batteries employ a gas diffusion electrode (GDE) which fulfills two functions. On the one hand, it supplies the cell with atmospheric oxygen, but keeps the electrolyte inside the cell. To this aim, the GDE contains a hydrophobic binder, repelling the aqueous electrolyte and enabling the coexistence of gas and liquid phases. On the other hand, the GDE reduces dissolved oxygen, providing hydroxide to the electrolyte. The cathode is filled with a non-noble catalyst, i.e., manganese oxide, to improve reaction kinetics at low costs. The specific surface area for oxygen reaction is enlarged by using a highly porous structure \cite{Neburchilov2010}.

\subsection{Reactions}
\label{sec:reakt}
\begin{figure}
	\includegraphics[width=\columnwidth]{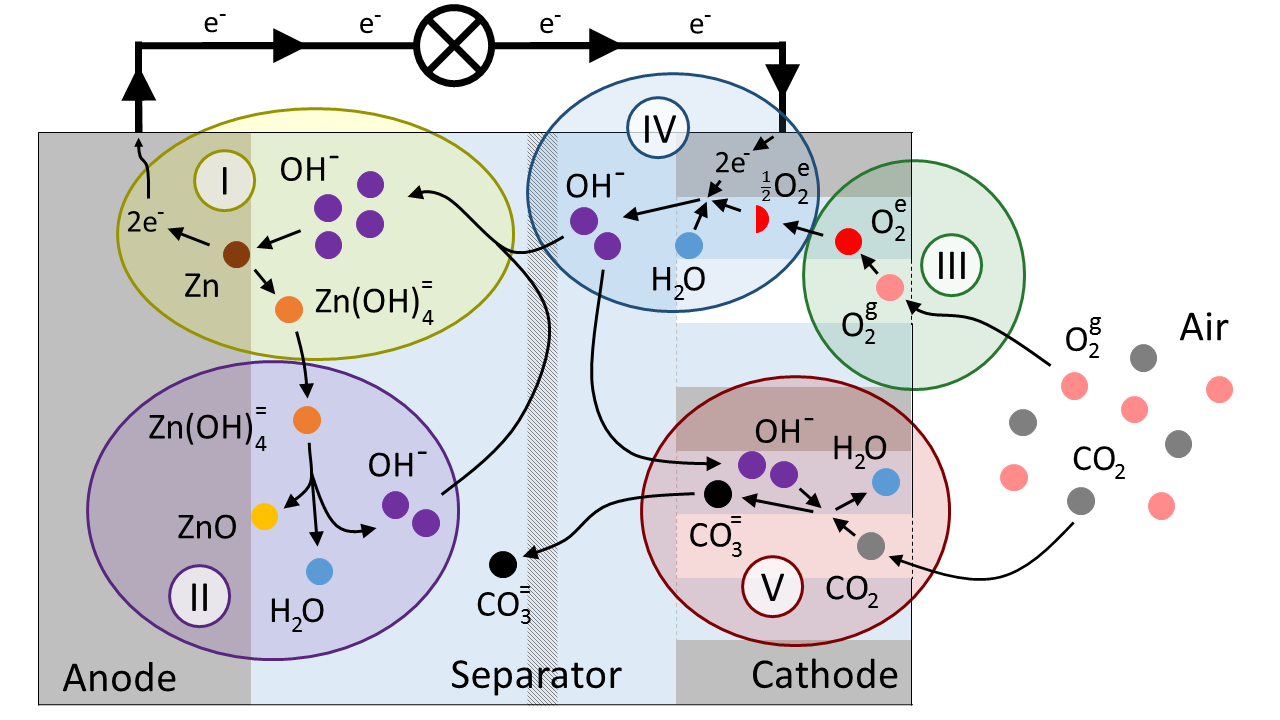}
\caption[Reaction scheme]{Reactions in the primary zinc-air cell: \ref{znox}) Zn dissolution, \ref{znoperc}) ZnO precipitation, \ref{o2abs}) Oxygen absorption into the electrolyte, \ref{o2red}) Oxygen reduction, \ref{co2abs}) Dissolution of atmospheric carbon dioxide and carbonate formation.}
\label{fig:react}
\end{figure}
The main reactions in a zinc-air battery are shown in Fig. \ref{fig:react}. During discharge the anodic Zn is oxidized. However, Zn does not directly transform into ZnO, but dissolves as zincate $\text{Zn(OH)}_x^{y-}$ into the electrolyte. The dominant type in the strongly alkaline electrolyte is $\text{Zn(OH)}_4^{=}$ \cite{Dirkse1954}, and we assume this is the only Zn species in the electrolyte. The chemical equation for the oxidation is
\begin{align*}
\chemtag
\label{znox}
\text{Zn}+4\text{ OH}^-\rightleftharpoons  \text{ Zn(OH)}^{=}_4 + 2 \text{ e}^-\,.
\end{align*}
The overall reaction can be divided into elementary first order reactions. Several reaction mechanisms are suggested with three \cite{Chang1984, Prentice1991} or four elementary reactions \cite{Bockris1972, Sunu1980}
\begin{align*}
\text{Zn}+\text{OH}^-&\rightleftharpoons \text{Zn(OH)}+ \text{e}^- \tag{\thechemequation.a} \\
\text{ZnOH}+\text{OH}^-&\rightleftharpoons \text{Zn(OH)}^-_2\tag{\thechemequation.b} \\
\text{Zn(OH)}^-_2 +\text{OH}^-&\rightleftharpoons \text{Zn(OH)}^-_3 + \text{e}^- \tag{\thechemequation.c} \\
\text{Zn(OH)}^-_3 +\text{OH}^-&\rightleftharpoons \text{Zn(OH)}^=_4\,. \tag{\thechemequation.d}
\end{align*}
In both cases oxidation of $\text{ Zn(OH)}^{-}_2$ is found to be rate limiting.

If the zincate concentration raises above its solubility limit, precipitation of ZnO becomes possible thermodynamically
\begin{align*}
\label{znoperc}
\chemtag
\text{Zn(OH)}^{=}_4 \rightleftharpoons  \text{ZnO}+ \text{H}_2\text{O}+2 \text{ OH}^-\,.
\end{align*}
The reaction takes place on the Zn surface and forms a porous ZnO layer, which retards the hydroxide supply of the anode. Thereby, it reduces the cell voltage and passivates the electrode once the layer is getting to thick \cite{Liu1981, Sunu1980}. If the overvoltage at the Zn surface becomes too large, ZnO type II forms as a compact ZnO layer and completely passivates the Zn. We omit ZnO type II in our model, since the cell voltage in our experiments does not allow its formation in the typical working domain $U> 1.1 \text{ V}$ \cite{Prentice1991}.

In the GDE at the gas-liquid phase boundary, atmospheric oxygen is dissolved \cite{Neidhardt2012} in the electrolyte 
\begin{align*}
\label{o2abs}
\chemtag
\text{O}_2^\text g\rightleftharpoons \text{O}_2^\text{ e}\,.
\end{align*}
Subsequently, dissolved oxygen is reduced to hydroxide at the active cathode surfaces
\begin{align*}
\label{o2red}
\chemtag
0.5 \text{ O}_2^\text e + \text{H}_2\text{O} + 2 \text{ e}^-\rightleftharpoons 2 \text{ OH}^-\,.
\end{align*}
The kinetics of oxygen reduction can be understood based on its elementary reaction steps \cite{eberle2014, Keith2010a} and depends on the employed catalyst \cite{Hansen2014}. 

The carbonate reaction is the major degradation process \cite{Drillet2001}. Atmospheric carbon dioxide dissolves and reacts to carbonate \cite{Ko1983}
\begin{align*}
\chemtag
	\label{co2abs}
	\text{CO}_2+2 \text{ OH}^- \rightleftharpoons    \text{CO}_3^{=} + \text{H$_2$O}\,.	
\end{align*}
In an alkaline medium the rate limiting reaction step is \cite{danckwerts1966}
\begin{align*}
 \text{CO}_2 + \text{OH}^- &\stackrel{k_\text{OH$^-$}}{\longrightarrow} \text{HCO$_3^-$}
 \label{co2absa}
\tag{\thechemequation.a}
\end{align*}
immediately followed by the reaction
\begin{align*}
	 \text{HCO$_3^-$} +\text{OH}^-  \longrightarrow  \text{CO$_3^{=}$} + \text{H$_2$O} \,.
	  \label{co2absc}
	 \tag{\thechemequation.b}
\end{align*}
The carbonate has various negative effects on the cell performance. The concentration of hydroxide, the main charge carrier, decreases. This leads to a loss of conductivity and enhances the passivation of the anode. Furthermore, carbonate inhibits both of the electrochemical reactions, because the decrease in hydroxide concentration reduces the solubilities of zincate and oxygen.

\section{Physical and Mathematical Model}
\label{sec:difmigmod} 
In this section, we introduce a thermodynamically consistent model for zinc-air cells. Our continuum model represents effects on a single dimension connecting anode, separator, and cathode. We start with a simple diffusion-migration model and successively add convection and reactions to it. First, we state a few central \emph{assumptions} which keep our model simple:
\begin{itemize}
\item We model an isothermal system since temperature variations are negligible in small zinc-air button cells.
\item The electrolyte is locally charge neutral because we are not interested in capacitive effects.
\item The electrolyte is strictly incompressible, i.e., its volume does not respond to pressure.
\item The partial pressures in the gas phase are constant because it is connected to the atmosphere and transport of gases is significantly faster than transport in the electrolyte.
\item No electrolyte is leaking out off the battery. Consequently, all electrolyte fluxes equal zero on the simulation domain boundaries.
\end{itemize}

In the following, we denote the solvent H$_2$O, the three kinds of anions OH$^-$, Zn(OH)$_4^{=}$, CO$_3^=$, and the cation K$^+$ with the indices $0$, $1$, $2$, $3$, and $+$, respectively.

\subsection{Electrolyte Diffusion and Migration}
\label{sec:transport}
We model diffusion and migration based on Latz et al. \cite{Latz2015, Zausch2010}. In this subsection, we discuss transport neglecting volume changes, chemical reactions, and convection. The latter means that we describe transport relative to the center of mass motion of the electrolyte. In the following subsections, we will step by step develop a realistic model with volume changes and chemical reactions, as well as convection. In the first part of this section, we present our model for pure electrolytes.

Dissolved oxygen can diffuse in the electrolyte
\begin{align}
\partial_tc_{\text{O}^2} =  \vec{\nabla}\cdot\left( D_{\text{O}^2} \vec{\nabla}c_{\text{O}^2} \right).
\end{align}
For the sake of clarity, we omit the dissolved oxygen below, knowing that it appears only in small amounts and does not influence the transport of the dominant species. According to Latz et al. \cite{Zausch2010}, the entropy production rate \nomenclature[R]{$\mathcal{R}$}{Entropy production rate} in polarizable systems in an external electromagnetic field in the isothermal case is
\begin{align}
\mathcal{R}= -\vec{j}\cdot \vec{\nabla}\Phi -  \sum_{i=1}^3\vec{N}_i \cdot \vec{\nabla} \mu_i\,.
\end{align}
Here\nomenclature[N]{$N_i$}{Particle flux density} $ \vec{N}_i$ denotes the particle flux density of species $i$, \(\vec{j}\) \nomenclature[j]{\(\vec{j}\)}{Current density} the current density, $\mu_i:=-\tilde\mu_0 (M_i+z_iM_+) M_0^{-1}+\tilde\mu_i+z_i\tilde\mu_+$ the \nomenclature[mu]{$\mu_i$}{Effective chemical potential}effective chemical potential and $\Phi$\nomenclature[Phi]{$\Phi$}{Electrical potential} the electrical potential, which holds $\vec{E}=-\vec{\nabla}\Phi$. The effective chemical potentials $\mu_i$ for the anions are valid in the center-of-mass frame assuming local charge neutrality.

The thermodynamical fluxes $\vec{j}$ and $\vec{N}_i$ fulfill the Onsager reciprocal relations, which we write compactly as
\begin{align}
	(-\vec{N}_1,-\vec{N}_2,	-\vec{N}_3,-\vec{\iota})^T
=
\mathcal{M}
\cdot
	(\vec{\nabla} \mu_1, \vec{\nabla} \mu_2, \vec{\nabla} \mu_3, \vec{\nabla} \phi)^T\,,
\end{align}
with the scaling $\vec{\iota} := \vec{j}F^{-1}$ and $\phi :=\Phi F$. In this scaling, the Onsager matrix $ \mathcal{M}$ is defined as
 \begin{align}
 \mathcal{M} := \mathcal{D} + \tilde\kappa \,\vec{\tau}\otimes\vec{\tau}\,,
\end{align}
with $\tilde \kappa := \kappa F^{-2}$, $\mathcal{D} := \text{diag}(\tilde D_1,\, \tilde D_2\,,\tilde D_3,\,0)$, \linebreak and $\vec{\tau} := (-\tau_1,\,-\tau_2\,,-\tau_3,\,1)^T$ with $\tau_i := t_i z_i^{-1}$. Here $t_i$, $\kappa$ and  $F$ denote the transference numbers\nomenclature[t]{$t_i$}{Transference number}, the electrolytic conductivity \nomenclature[kappa]{$\kappa$}{Electrolytic conductivity} and the Faraday constant\nomenclature[F]{\(F\)}{Faraday constant}, respectively.

  The Onsager matrix has to be positive-semidefinite, since the entropy production rate $\mathcal{R}$ is always non-negative in a physical system. This is obviously fulfilled, if $\tilde D_i\geq0$ and $\tilde \kappa\geq0$, due to the simple calculation
\begin{align}
	\vec{x}^T \mathcal{M}\, \vec{x} =\sum_{i=1}^n \tilde D_i x_i^2 + \tilde\kappa\, \left(\vec{x}\cdot\vec{\tau}\right)\,.
\end{align}
The consequences of the reciprocal relations for the thermodynamical fluxes are more apparent in the standard notation. We express these equations in terms of concentrations, which is more convenient. Assuming that the chemical potential $\mu_i\equiv\mu_i(c_i)$ of any species depends on the corresponding concentration only, we find
 \begin{align}
	\label{ionenfl}
	\vec{N}_i&= - D_i \vec{\nabla}c_i - \frac{t_i}{z_iF}\vec{j}\,,
	\\
	\label{stromfl}
	\vec{j} &=  -\kappa \vec{\nabla}\Phi + \frac{\kappa}{F}  \sum_{i=1}^3\frac{t_i}{z_i} \left(\frac{\partial \mu_i}{\partial c_i}\right) \vec{\nabla}c_i \,.
\end{align}
with $D_i:=\tilde D_i  (\frac{\partial \mu_i}{\partial c_i}) $ denoting the \nomenclature[D]{$D_i$}{Diffusion coefficient}diffusion coefficients.

In the absence of reactions, concentration and charge density are conserved. Taking the local charge neutrality of the electrolyte into account, they satisfy the continuity equations
\begin{align}
	\label{teifl}
	 \partial_tc_i=-\vec{\nabla}\cdot \vec{N}_i && \text{and} && 0=-\vec{\nabla}\cdot \vec{j}\,.
\end{align}
Combining these relations with Eqs. \ref{ionenfl} and \ref{stromfl}, we finally find the transport equations
\begin{align}
	\label{tpgl}
	\partial_tc_i& =  \vec{\nabla}\cdot\left( D_i \vec{\nabla}c_i \right)+ \vec{\nabla}\cdot\left(\frac{t_i}{z_iF}\vec{j}\right)
	\\
	\label{stromgl}
	0 &= \vec{\nabla}\cdot\left(\kappa \vec{\nabla} \Phi\right) - \vec{\nabla}\cdot\left(\frac{\kappa}{F}  \sum_{i=1}^3\frac{t_i}{z_i} 			\left(\frac{\partial \mu_i}{\partial c_i}\right) \vec{\nabla}c_i\right) .
\end{align}

In the second part of this section, we consider the electrodes. Therefore, we now allow three coexisting phases: gas, liquid, and solid.

Taking the porous electrode into account changes the model in two ways: fluxes in porous media differ from unhindered fluxes in pure liquids and the transport equations are only applied to the electrolyte volume.

We define the volume fraction of each phase as $ \epsilon_{i}:=\frac{V_{i}}{V_\text{total}}$\nomenclature[epsilon]{$ \epsilon_{i}$}{Volume fraction}. This definition obviously yields
\begin{align}
	\label{sumeps}
	1 =  \epsilon_\text{s}+ \epsilon_\text{e}+ \epsilon_\text{g}\,,
\end{align} 
where the indices s, g, and e denote the solid phase, the gas phase, and the electrolyte, respectively. In agreement with Ref. \cite{Newman2004}, we model the effects of porosity and tortuosity on the fluxes via the factor $\epsilon_\text{e}^\beta$, with the Bruggeman coefficient $\beta=1.5$\nomenclature[beta]{$\beta$}{Bruggeman coefficient}. 

The effective flux density equations are (see eq. \ref{ionenfl}, \ref{stromfl})
 \begin{align}
	\label{ionenfleff}
	\vec{N}_i^\text{eff}&= - \epsilon_\text{e}^\beta D_i \vec{\nabla}c_i - \epsilon_\text{e}^\beta\frac{t_i}{z_iF}\vec{j}\,,
	\\
	\label{stromfleff}
	\vec{j}^\text{eff} &=  -\epsilon_\text{e}^\beta\kappa \vec{\nabla}\Phi + \frac{\epsilon_\text{e}^\beta\kappa}{F}  \sum_{i=1}^3\frac{t_i}{z_i} \left(\frac{\partial \mu_i}{\partial c_i}\right) \vec{\nabla}c_i \,.
\end{align}

\subsection{Reactions in Porous Electrodes}
\label{sec:reaktmod}
In this section, we incorporate the reactions described in Sec. \ref{sec:reakt} into our model. Then particle flux density and current density are not conserved quantities. In our macro-homogeneous approach, reactions appear as species-related source terms $S_i$ \nomenclature[S]{$S_i$}{Species-related source term} (see eq. \ref{sourcesum}) in the transport equations. Thus, the continuity Eqs. \ref{teifl} become
\begin{align}
	\label{teiflreact}
	 \partial_t( \epsilon_\text{e} c_i ) &= -\vec{\nabla}\cdot \vec{N}_i^\text{eff}+S_i\quad \text{and}
	\\
	\label{teifl2react}
  			0 &= -\vec{\nabla}\cdot \vec{j}^\text{eff}- \sum_{i=1}^3z_i F S_i  \,.
\end{align}
We find the macro-homogeneous transport equations with the definition of the effective fluxes \ref{ionenfleff} and \ref{stromfleff}
\begin{align}
	\label{tpglreact}
	\partial_t( \epsilon_\text{e}c_i )& =  \vec{\nabla}\cdot\left(\epsilon_\text{e}^\beta D_i \vec{\nabla}c_i \right)								+\vec{\nabla}\cdot\left(\epsilon_\text{e}^\beta\frac{t_i}{z_iF}\vec{j}\right)+S_i
	\\
	\label{stromglreact}
	\begin{split}
	0 &=  - \sum_{i=1}^3z_i F S_i +\vec{\nabla}\cdot\left(\epsilon_\text{e}^\beta\kappa \nabla\Phi\right)
	\\
	& \hspace{4em} -\vec{\nabla}\cdot\left(\frac{\epsilon_\text{e}^\beta \kappa}{F}  \sum_{i=1}^3\frac{t_i}{z_i} 			\left(\frac{\partial \mu_i}{\partial c_i}\right) \vec{\nabla}c_i\right)\,.
	\end{split}
\end{align}
The species-related source terms depend on the reaction-specific source terms $s_j$
\begin{align}
\label{sourcesum}
  S_i = \sum_{j\in\mathcal{J}} s_j \nu_{ij}\, ,
\end{align}
where $\mathcal{J}=\{\text{I},\dots,\text{V}\}$ denotes the set of all reaction indices and 
\nomenclature[nu]{$\nu_{ij}$}{Stoichiometric index}$\nu_{ij}$ the stoichiometric index of species $i$ in reaction $j$. The reaction-specific source terms $s_j$ are discussed in the following. Generally, the source terms $s_i = A_i\cdot j_i$ are the products of the surface-related reaction rates $j_i$ \nomenclature[j]{$j_i$}{Surface-related reaction rate} and the specific surface areas $A_i$\nomenclature[A]{$A_i$}{Specific surface area}.

\subsubsection{Zn oxidation / dissolution}
To model the surface-related reaction rate of first order electrochemical reactions, a thermodynamically consistent Butler-Volmer approach is applied \cite{Latz2013, Bazant2012}.
\begin{align}
j=j_0\left[\exp\left(\alpha\frac{ z F}{RT}\eta\right)-\exp\left(-(1-\alpha) \frac{z F}{RT}\eta\right)\right]\,,
\end{align}
with the exchange current density \nomenclature[j]{$j_0$}{Exchange current density}
\begin{align}
	j_0 := k \left(\frac{c_\text{O}}{c_\text{std}}\right)^{(1-\alpha)}\left(\frac{c_\text{R}}{c_\text{std}}\right)^{\alpha}
\end{align}
and the \nomenclature[eta]{$\eta$}{Activation overpotential} activation overpotential
\begin{align}	
	\eta= \eta^0+\frac{ RT}{zF}\ln\left(\frac{c_\text{O}}{c_\text R}\right)\,.
\end{align}
Here $R$\nomenclature[R]{$R$}{Universal gas constant}, $\alpha$\nomenclature[alpha]{$\alpha$}{Symmetry factor}, $z$\nomenclature[z]{$z$}{Number of exchanged electrons per reaction}, $c_\text{O}$, and $c_\text{R}$ denote the universal gas constant, the symmetry factor, the number of exchanged electrons, and the concentration of the oxidizing and the reducing agents, respectively. The second term on the right hand side takes the chemical potential differences into account, which are caused by the species concentrations in the electrolyte.

Even though the Zn oxidation is not an elementary one-electron reaction (see Reaction \ref{znox} in Sec. \ref{sec:reakt}), complex rate expressions based on the rate determining electron transfer exist \cite{Sunu1980, Bockris1972}. As, however, diffusion through the ZnO layer is limiting Zn dissolution, we can employ the simpler consistent global rate expression \cite{Deiss2002, Schroder2014}
\begin{align}
\label{butvola}
j_\text{\ref{znox}}=2 k_\text{\ref{znox}} \sqrt{\frac{c_\text{s,OH$^-$}^4c_\text{Zn(OH)$^{=}_4$}^{ }}{c_\text{std}^5}}\sinh\left(\frac{F}{RT}\eta_\text{a}\right)\,,
\end{align}
where we choose the symmetry factor $\alpha=0.5$. Here $k_\text{\ref{znox}}$\nomenclature[k]{$k_i$}{Kinetic coefficient} denotes the kinetic coefficient and $c_{\text{s,OH}^-}$ the hydroxide concentration at the Zn surface, in contrast to the hydroxide bulk concentration $c_{\text{OH}^-}=c_{\text{b,OH}^-}$. The  overpotential yields
\begin{align}	
\label{overpotano}
	\eta_\text{a}= \Delta \phi_\text{a} - \Delta \phi_\text{a}^0 +\frac{ RT}{zF}\ln\left(\frac{c_\text{s,OH$^-$}^4}{c_\text{Zn(OH)$^{=}_4$}^{ }  c_\text{std}^3}\right)\,.
\end{align}
Thereby, $\Delta\phi_\text{a}:=\phi_\text{a}-\phi_\text{e}$ is the potential difference between anode and electrolyte and $\Delta \phi^0_\text{a}$ is the open circuit half-cell potential at standard concentrations. 

For determining the specific surface area $A_\text{\ref{znox}}$\nomenclature[A]{$A_i$}{Specific surface area} of the anode, we assume that the Zn electrode consists of spherical particles with radius $r_\text{Zn}$\nomenclature[ri]{$r_i$}{Radius}. Thus, the constant density of spherical Zn particles 
\begin{align}
N_\text{Zn} = \frac{3\epsilon_{Zn}^0}{4 \pi}\left(r_\text{Zn}^0\right)^{-3}
\end{align}
depends on initial volume fraction $\epsilon_{Zn}^0$ and radius $r_\text{Zn}^0$. At each time step and position, we calculate Zn radius and specific surface area according to
\begin{align}
\label{znr}
r_\text{Zn}=\left(\frac{3\epsilon_\text{Zn}}{4 \pi N_\text{Zn}}\right)^{\frac{1}{3}},\hspace{0.5cm}
A_\text{\ref{znox}} = 4\pi N_\text{Zn} r_\text{Zn}^2\,.
\end{align}
The ZnO layer formed around dissolved Zn particles was imaged by Shao-Horn \cite{Horn2003,Szpak1979}. First, ZnO type I forms a porous shell filled with electrolyte and the remaining Zn particle. We assume that the ZnO film forms uniformly with a constant porosity $\epsilon_\text{f}$ on each of the Zn particles in a certain control volume. The film thickness is $\delta_\text{Zn}:= r_\text{ZnO}-r_\text{Zn}^0$\nomenclature[delta]{$\delta$}{Thickness of ZnO film} with the constant inner radius $r_\text{Zn}^0$ and the growing outer radius
\begin{equation}
r_\text{ZnO}=r_\text{Zn}^0\left(1+\frac{1}{1-\epsilon_\text{f}}\frac{\epsilon_\text{ZnO}}{\epsilon_\text{Zn}^0}\right)^{\frac{1}{3}}\,.
\end{equation}
The hydroxide concentration $c_{\text{s,OH}^-}$ at the Zn surface is limited by diffusion through the porous film and by hydroxide consumption due to Zn oxidation $s_\text{\ref{znox}}$ at the surface \cite{Bockris1972,Liu1981}. Hydroxide transport is described by spherical diffusion 
\begin{align}
\label{portrans}
	\frac{4 s_\text{\ref{znox}}}{A_\text{\ref{znox}}^0}
		=  \epsilon_\text{f}^{3.5} D_{\text{OH}^-}  \frac{c_{\text{b,OH}^-} - c_{\text{s,OH}^-}}{\delta_\text{Zn}}\frac{r_\text{ZnO}}{r_\text{Zn}^0}\,.
\end{align}
driven by the concentration gradient between the bulk $c_{\text{b,OH}^-}$ and surface $c_{\text{s,OH}^-}$ concentration. We increase the Bruggemann coefficient to $3.5$ here, in order to simulate the diffusion limitation proven experimentally. This value is realistic for compact materials \cite{Nyman2008,Single2016}.

\subsubsection{Oxygen reduction}
We model the rate of oxygen reduction (see Reaction \ref{o2red}) via the symmetric Butler-Volmer approach \cite{Cao2003} 
\begin{align}
\label {redrate}
	j_\text{\ref{o2red}} = -2 k_\text{\ref{o2red}} \frac{c_\text{OH$^-$}}{c_\text{std}}
	\sqrt[4]{\frac{c_{\text{O}_2}^{ }}{c_\text{std}^{\text{O}_2}}}
	\sinh	\left(\frac{F}{RT}\eta_\text{c}\right)\,,
\end{align}
with the activation overpotential
\begin{align}
\label{catpot}	
	\eta_\text{c}= \Delta \phi_\text{c} - \Delta \phi_\text{c}^0 +\frac{ RT}{zF}\ln\left(\sqrt{\frac{c^\text{O$_2$}_\text{std}}{c_\text{O$_2$}^{\rule{0em}{0.3em} }}} \frac{c_\text{OH$^-$}^2}{c_\text{std}^2}\right)\,.
\end{align}
The oxygen reduction overpotential dominates the cell overpotential, but remains almost constant during discharge. This justifies our approach to make use of simple, but consistent global reaction kinetics. More complex rate expressions are discussed elsewhere \cite{eberle2014}.

The oxygen reduction does not change the surface of the GDE. Our simulations show that almost no ZnO precipitates in the cathode. Consequently, we assume that the specific surface $A_\text{\ref{o2red}}$ remains constant during discharge. 

\subsubsection{ZnO nucleation and growth} 
Thermodynamics allows ZnO to grow for concentrations above the solubility limit $c_\text{sat}$ of zincate (see Reaction \ref{znoperc}). However, nucleation requires greater concentrations \cite{Horstmann2013} and supersaturation ratios up to $s:=c_{\text{Zn(OH)}_4^{=}}/c_\text{sat}\approx 4$ are reported for Zn anodes \cite{Liu1981}. Precipitation is typically diffusion limited for reactions with large supersaturation ratios $s$ \cite{Sunu1980,Horstmann2013}, yielding 
\begin{align}
\label{znopercreact}
j_\text{\ref{znoperc}} = \epsilon_\text{f}^{3.5}D_{\text{Zn(OH)}_4^{=}}\frac{c_{\text{Zn(OH)}_4^{=}} - c_\text{sat} }{\delta_\text{ZnO}} \,
\end{align}
with the diffusion layer thickness $\delta_\text{ZnO}$. Here we apply the same Bruggemann factor as in Eq. \ref{portrans}. The specific surface area for ZnO precipitation $A_\text{\ref{znoperc}}$ depends on nucleation and growth of ZnO particles. This process can be described with classical nucleation theory \cite{Horstmann2013}. We apply a more phenomenological approach to keep the model numerically simple. Nucleation occurs abruptly in our model if the concentration exceeds a critical supersaturation $c_\text{crit}$, an additional parameter. The specific surface area is
\begin{align}
\label{znosurf}
	 A_\text{\ref{znoperc}} = \begin{cases}
	 							\makebox[6em][l]{ $4\pi N_\text{Zn} r_\text{ZnO}^2$}\makebox[12em]{$c_{\text{Zn(OH)}_4^{=}}>c_\text{crit} \,\,\vee \,\,\epsilon_\text{ZnO} > \epsilon_\text{ZnO}^0$} 
	 							\\
	 							\makebox[6em][l]{ 0} \makebox[12em]{else.}
							\end{cases}
\end{align}
To avoid a discontinuity, we linearly ramp up the specific surface area until $100$ ZnO monolayers are deposited. 

\subsubsection{Oxygen dissolution}
The solubility of oxygen in water (see Reaction \ref{o2abs}) depends linearly on the partial oxygen pressure via Henry's Law \cite{Neidhardt2012,Horstmann2013}
\begin{align}
\label{henry}
c^*_{\text{O}_2} = 10^{-K^\text{s}_{\text{O}_2}} H^{\text{c,p}}_{\text{O}_2}p_{\text{O}_2} \,,
\end{align}
where $H^{\text{c,p}}$\nomenclature[Hcp]{$H^{\text{c,p}}$}{Henry constant} is Henry's constant and $p_{\text{O}_2}^{}$\nomenclature[p]{$p_i$}{Partial pressure} the partial oxygen pressure. The dependence of solubility on salt concentration, denoted salting out, is described with the Sechenov constant $K^\text{s}$ \cite{Weisenberger1996}. 

The kinetics of oxygen dissolution is given by the Hertz-Knudsen equation \cite{Horstmann2013,Eames1997}
\begin{align}
	\label{o2absreac}
 	j_\text{\ref{o2abs}} = \frac{p_{\text{O}_2}\xi}{ c^*_{\text{O}_2}(2\pi M_{\text{O}_2}RT)^{0.5} } \left(c^*_{\text{O}_2}- c^{}_{\text{O}_2} \right)\,.
\end{align}
where $\xi$ denotes the ratio of dissolved molecules to molecules hitting the gas-liquid phase boundary. The specific surface area $A_\text{\ref{o2abs}}$ corresponds to the gas-liquid phase boundary and is assumed constant during the discharge process.

\subsubsection{Carbon dioxide absorption}
Upon absorption, carbon dioxide immediately reacts and forms carbonate (see Reaction \ref{co2abs}). Due to its high rate, this reaction takes place in a small layer at the gas-liquid phase boundary thinner than the resolution of our 1D model. Therefore, we include a simplified macroscopic pseudo first-order reaction rate in our cell model. In the following, we derive it from a microscopic diffusion-reaction model \cite{Cents2003, Cents2005, Danckwerts1950}.

We calculate the concentration of dissolved carbon dioxide in one dimension ($y\in [0,\infty)$) perpendicular to the phase boundary at $y=0$. Diffusion determines its transport since the pressure gradient in the thin surface layer is negligible. Hence, the carbon dioxide concentration fulfills the simple diffusion-reaction equation \cite{Danckwerts1950}
\begin{align}
\label{co2pdgl1}
\partial_t c^{}_{ \text{CO}_2 }(y,t) &= D_{ \text{CO}_2} \partial_y^2 c^{}_{ \text{CO}_2}(y,t)  - s_{\text{CO}_3^{=}}(y,t),
\end{align}
with the rate $s_{\text{CO}_3^{=}}$ of Reaction \ref{co2abs} (see below).

Let us assume that the carbon dioxide concentration remains in equilibrium at the phase boundary. We estimate the solubility $c^*_{\text{CO}_2}$ for the dissolution of carbon dioxide with Henry's law and find the boundary condition 
\begin{align}
\label{co2pdgl2}
 c^{}_{ \text{CO}_2 }(0,t) &= c^*_{\text{CO}_2} = 10^{-K^\text{s}_{\text{CO}_2}} H^{\text{c,p}}_{\text{CO}_2}p^{}_{\text{CO}_2} \,.
\end{align}

The rate of the microscopic Reaction \ref{co2abs} depends linearly on the deviation of the concentration from equilibrium, according to Danckwerts et al. \cite{danckwerts1966}
\begin{align}
s^{}_{\text{CO}_3^{=}} = k^{}_{\text{OH}^-} c^{}_{\text{OH}^-}(c^{}_{ \text{CO}_2}-c_{ \text{CO}_2}^\text{eq})\,c^{-2}_{\text{std}}\,.
\end{align}
Here $k_{\text{OH}^-}$ denotes the kinetic constant of the rate determining step (see Reaction \ref{co2absa}) for carbon dioxide absorption in alkaline media. The equilibrium concentration $c_{\text{CO}_2}^\text{eq}\approx 0$ is negligibly small. In the thin film, we assume a constant hydroxide concentration. The simplified local reaction rate yields
\begin{align}
\label{co2reactloc}
s^{}_{\text{CO}_3^{=}} =k^{}_{\text{OH}^-} c^{}_{\text{OH}^-} c^{}_{ \text{CO}_2} c^{-2}_{\text{std}}\,.
\end{align}
On macroscopic time scales, reaction and diffusion through the thin surface layer are fast. Thus, the concentration profile is stationary $c_{ \text{CO}_2 }(0,t)=0$. This simplifies the partial differential equation \ref{co2pdgl1} to the ordinary differential equation
\begin{align}
\label{rdgl}
D_{ \text{CO}_3^{=}} \partial_x^2c^{}_{ \text{CO}_2}(y)  = k^{}_{\text{OH}^-} c^{}_{\text{OH}^-} c^{}_{ \text{CO}_2}(y)\,c^{-2}_{\text{std}}\,. 
\end{align}
We solve for the concentration profile
\begin{align}
	 c^{}_{ \text{CO}_2 }(y) = c^*_{\text{CO}_2} \exp\left(-\sqrt{\frac{k_{\text{OH}^-} c^{}_{\text{OH}^-}}{D_{ \text{CO}_2} c^{2}_{\text{std}}}}\,y\right)\,.
\end{align}
Next, we evaluate the macroscopic reaction rate by integrating the microscopic, local reaction rate $s_{\text{CO}_3^{=}}(y)$ and get
\begin{align}
  	j_\text{\ref{co2abs}} \quad=& \quad  \int_0^\infty  s_{\text{CO}_3^{=}}(y) \,dy
  	\\
  	\quad=& \quad c^*_{\text{CO}_2}\sqrt{k^{}_{\text{OH}^-} c^{}_{\text{OH}^-}D_{ \text{CO}_2} } c^{-1}_{\text{std}}\,.
\end{align}
The surface area for oxygen and carbon dioxide absorption $A_\text{\ref{co2abs}} = A_\text{\ref{o2abs}}$ is the gas-liquid phase boundary surface.

\subsection{Electrolyte Convection}
\label{sec:konvmod}
Reactions can change the volume available for electrolyte (see Sec. \ref{sec:volbil}) and the electrolyte composition (see Sec. \ref{sec:reaktmod}). For incompressible electrolytes, such volume changes lead to convection. In this section we add convection to the transport theory described in Sec. \ref{sec:transport} determining transport relative to the center of mass. The convective velocity $\vec v$ is defined via the flux of the center of mass motion $\rho := \sum_i M_i c_i$ \cite{Latz2015} 
\begin{align}
\label{tprho}
  	\partial_t  (\epsilon_\text{e}\rho)= \vec{\nabla}\cdot\left(\epsilon_\text{e}^\beta\rho\vec{v}_\text{e}\right)+\sum_i M_i S_i \,.
\end{align}
We employ this transport equation to calculate the concentration of water in our model.

The convective flux density of each species in the electrolyte is $\vec{N}_i^\text{conv,eff}= \epsilon_\text{e}^\beta c_i \vec{v_\text{e}}$. Above we discuss diffusion and migration relative to the center of mass. Therefore, the convective flux density is added to the transport Eqs. \ref{tpglreact}
\begin{align}
	\label{tpglkonv2}
	\partial_t( \epsilon_\text{e}c_i )& =  \vec{\nabla}\cdot\left(\epsilon_\text{e}^\beta D_i \vec{\nabla}c_i \right)								+\vec{\nabla}\cdot\left(\epsilon_\text{e}^\beta\frac{t_i}{z_iF}\vec{j}\right)
	\notag
	\\
	&\qquad+\vec{\nabla}\cdot\left(\epsilon_\text{e}^\beta c_i\vec{v}_\text{e}\right)+S_i \,.
\end{align}
Note that convection has no influence on the current density because the electrolyte is locally charge neutral.

Next, we describe how the convective velocity depends on electrolyte composition. Our Ansatz is that the convective velocity is such that the electrolyte equation of state remains fulfilled \cite{Horstmann2013, Danner2014}. The electrolyte equation of state can be expressed in terms of volumes. This is non-trivial  because in general $V_\text{solution}\neq V_\text{solute}+V_\text{solvent}$. The volume change of the solution, caused by adding one more particle of species $i$, is denoted  partial molar volume \nomenclature[V_i]{$\bar V_i$}{Partial molar volume}$\bar V_i$ of species $i$. At constant pressure and temperature it still depends on the composition of the solution \cite{Groot}. Since the volume is an extensive property, the equation of state is
\begin{align}
\label{volbed}
	 1=  \sum_{i=1}^kc_i\bar V_i(c_1,\dots,c_k)\,,
\end{align}
where we parametrize the partial molar volumes $\bar V_i$ as a function of electrolyte composition in this paper. Together with Eq. \ref{tpglkonv2}, we find the following equation for the convective velocity
\begin{align}
\vec{\nabla}\cdot\left(\epsilon_\text{e}^\beta \vec{v}_\text{e}\right) = \partial_t \epsilon_\text{e}-\sum_{i=1}^k \bar V_i \left[ S_i +
\vec{\nabla}\cdot\left(\epsilon_\text{e}^\beta D_i \vec{\nabla}c_i \right)								+\vec{\nabla}\cdot\left(\epsilon_\text{e}^\beta\frac{t_i}{z_iF}\vec{j} \right)\right]\,.
\end{align}
Continuum models of gas diffusion electrodes in fuel cells and metal-air batteries \cite{Horstmann2013, Jahnke2016} use Darcy's law to connect electrolyte velocity \nomenclature[ve]{$\vec{v}_\text{e}$}{Velocity electrolyte}$\vec{v}_\text{e}$ and pressure $ p_\text{e}$ in porous media 
\begin{align}
	\label{darcylaw}
 	\vec{v}_\text{e} = -\frac{B_\text{e}}{\eta_\text{e}} \vec{\nabla} p_\text{e}\,.
\end{align}
Here $B_\text{e}$\nomenclature[Be]{$B_\text{e}$}{Permeability} denotes the permeability of the electrodes with respect to the electrolyte and $\eta_\text{e}$ \nomenclature[etae]{$\eta_\text{e}$}{Dynamic viscosity} the dynamic viscosity of the electrolyte.

\subsection{Solid and Gas Phases}
\label{sec:volbil}
In this subsection, we describe volume changes due to transport and reactions. The dynamics of the volume fractions of solid phases is determined by the appropriate source terms \cite{Neidhardt2012}
\begin{align}
\label{volsol}
 	\partial_t\epsilon_\text{Zn}=\bar V_\text{Zn} S_\text{Zn}\,& &\text{and}&& \partial_t\epsilon_\text{ZnO}=\bar V_\text{ZnO} S_\text{ZnO}\,
\end{align}
with the constant molar volumes $\bar V_i$.

In the real button cell, the gas phase is present in a compact void space under the anode lid (see Sec. \ref{sec:aufbau}) and in the gas diffusion electrode. In our model, we consider this void space to be evenly distributed throughout the anode and the separator and keep the model numerically simple. Due to its large kinematic viscosity, convection of gas is two orders of magnitude faster than convection of electrolyte at the same pressure gradient. Therefore, we assume that the partial pressures $p_{\text{CO}_2}$ and $p_{\text{O}_2}$ as well as the overall pressure $p_\text{g}$ remain constant throughout the cell.

As frequently done for gas diffusion electrodes \cite{Horstmann2013, Danner2014}, our model relies on pressure saturation curves which can either be measured or calculated {with 3D Lattice-Boltzmann simulations \cite{Danner2016}}. It is a complex task to lay out the gas diffusion electrode such that it contains an even mixture of electrolyte and gas phase. Commercial button cells, however, are well-designed and the specific form of the Leverett J-function does hardly influence our simulation results. Let the saturation $\tilde s$ \nomenclature[s]{$\tilde s$}{Saturation} of the porous media be the ratio of the electrolyte volume to the void space
$\tilde s:={V_\text{e}}/{(V^0_\text{e}+V^0_\text{g})}={ \epsilon_\text{e}}/{ (\epsilon^0_\text{e}+ \epsilon^0_\text{g})}$. Then the saturation is determined by the electrolyte pressure (see Eq. \ref{darcylaw}) via the Leverett J-function \cite{Horstmann2013, Danner2014}
\begin{align}
 	J(\tilde s) =\sqrt{\frac{B_\text{e}}{\epsilon_\text{s}\sigma^2}}p_\text{c} :=\sqrt{\frac{B_\text{e}}{\epsilon_\text{s}\sigma^2}}
 	(p_\text{e}-p_\text{g})\,.
\end{align}
Here $p_\text{e}$, $p_\text{g}$, $p_\text{c}$ denote the pressure in the electrolyte, the gas phase, the capillary pressure, respectively. $\sigma$\nomenclature[sigma]{$\sigma$}{Surface tension} is the surface tension between electrolyte and GDE and $B_\text{e}$ is the GDE permeability for electrolyte.

\subsection{Galvanostatic Condition}
The external current density $i_\text{cell}$ must always match the density of exchanged electrons in the electrochemical reactions in each of the electrodes \cite{Zausch2010}. For the cathode this yields
 \begin{align}
 \label{galvo}
 	i_\text{cell} = \int_{V_\text{c}} z_\text{\ref{o2red}} F s_\text{\ref{o2red}} \,dx\,.
\end{align}
Since the electrolyte is modeled charge neutral, it is sufficient to consider this constraint in a single electrode (see Eq. \ref{stromglreact}).

\section{Parameterization and Computational Details}
\label{sec:parameters}
\subsection{Parameterization}
We model the Varta PowerOne hearing aid coin cell battery PR44 type p675. Therefore, the parameters represent this battery type. Decades ago, thermodynamics \cite{Lange1999,Weisenberger1996,Sander2014,Sunu1978,Siu1997,moller2003} and ionic transport \cite{Davis1967,May1978,bhatia1968,Zeebe2011,Cents2003,Sunu1980,Liu1981b,See1997,Lange1999, Liu1981b,Newman2004,Siu1997,Danner2015,Group1962} in the aqueous alkaline electrolyte (32 weight percent KOH) were accurately studied with experiments. We discuss the parameters in the Supplementary Materials A. Our thermodynamic parameters and transport parameters are based on the extensive literature. In contrast, the reaction kinetics are not known with sufficient accuracy. Therefore, we choose to adjust them such that the simulated discharge curves match the measured ones. Nevertheless, we make sure that the reaction parameters are reasonable by comparing to the literature data \cite{Manke2007,Liu1981,Deiss2002,Sunu1980,Gyenge2012,Horstmann2013,pohorecki1988}. We want to highlight that the qualitative features of our simulation results are robust against variations of the kinetic parameters.

\subsection{Computational Details}
For the simulations, we implement our model in Matlab. The finite volume method is used for space discretization \cite{popov2010}. Time evolution is  performed by the implicit, Matlab built-in solver ode15i.

\section{Experimental Setup and Procedure}
\label{sec:setup}
The electrochemical experiments were carried out with commercial Varta PowerOne hearing aid batteries PR44 Type p675 on a multichannel modular potentiostat/galvanostat VMP3 from Bio-Logic Science Instruments (France). According to the IEC 60086-2 norm, the seal of the cells was removed 10 minutes before starting each experiment, in order to activate the battery. 

Afterwards, two different kinds of test were performed on the commercial zinc-air cells: 
\begin{itemize}
	\item \emph{Galvanostatic discharge} After recording the open circuit voltage (OCV) for 30 seconds, the cells were discharged by applying a constant current ranging from 25 to 125 Am$^{-2}$. The voltage was monitored over time, until it reached the value of 0.9 V, which was selected as the end of discharge cut-off.
	\item \emph{Lifetime analysis} Firstly, in order to reach the voltage plateau, the cells were subject to a galvanostatic discharge step, whose length and current density was selected to be either 5h at 100 Am$^{-2}$ or 10h at 50 Am$^{-2}$. Afterwards, the cells were left to relax and the OCV recorded for 24h. After such rest period the cells were partially discharged with a constant current pulse of 100 Am$^{-2}$ or 50 Am$^{-2}$ for 10 minutes. The cell voltage at the end of each pulse was used to monitor the aging of the cell (see Figure B.1 in the Supplementary Materials). Such OCV-pulse pattern was repeated for several days, until the voltage at the end of the pulse dropped to the cut-off value of 0.9 V.
\end{itemize}
The tests were performed at room temperature and atmosphere if not stated otherwise. Therefore, small fluctuations in the discharge voltage profiles can be addressed to uncontrollable environmental changes in the laboratory over the experiment time-span.

In order to ensure the reproducibility of the experimental results, each kind of test was repeated at least once. Despite small variation due to uncontrollable factors (e.g., air flow in the laboratory, temperature fluctuation, and eventual differences among the cells coming from the factory), the qualitative features necessary to validate the modeling was always observed. For sake of brevity, we report the most representative measurements only.
\section{Galvanostatic Discharge}
\label{sec:discharge}
In this section, we discuss the discharge of the zinc-air button cell at various currents. Our simulations allow to study internal variables like ion concentrations and phase distributions which are not directly accessible experimentally. Therefore, we interpret the experimental and theoretical discharge curves by analyzing the simulated internal variables in parallel. The procedure for experiment and simulation is described in detail in Sec. \ref{sec:setup}.

\subsection{Experiment}
\label{sec:expdischarge}
\begin{figure}
	\centering
	\includegraphics[width=\columnwidth]{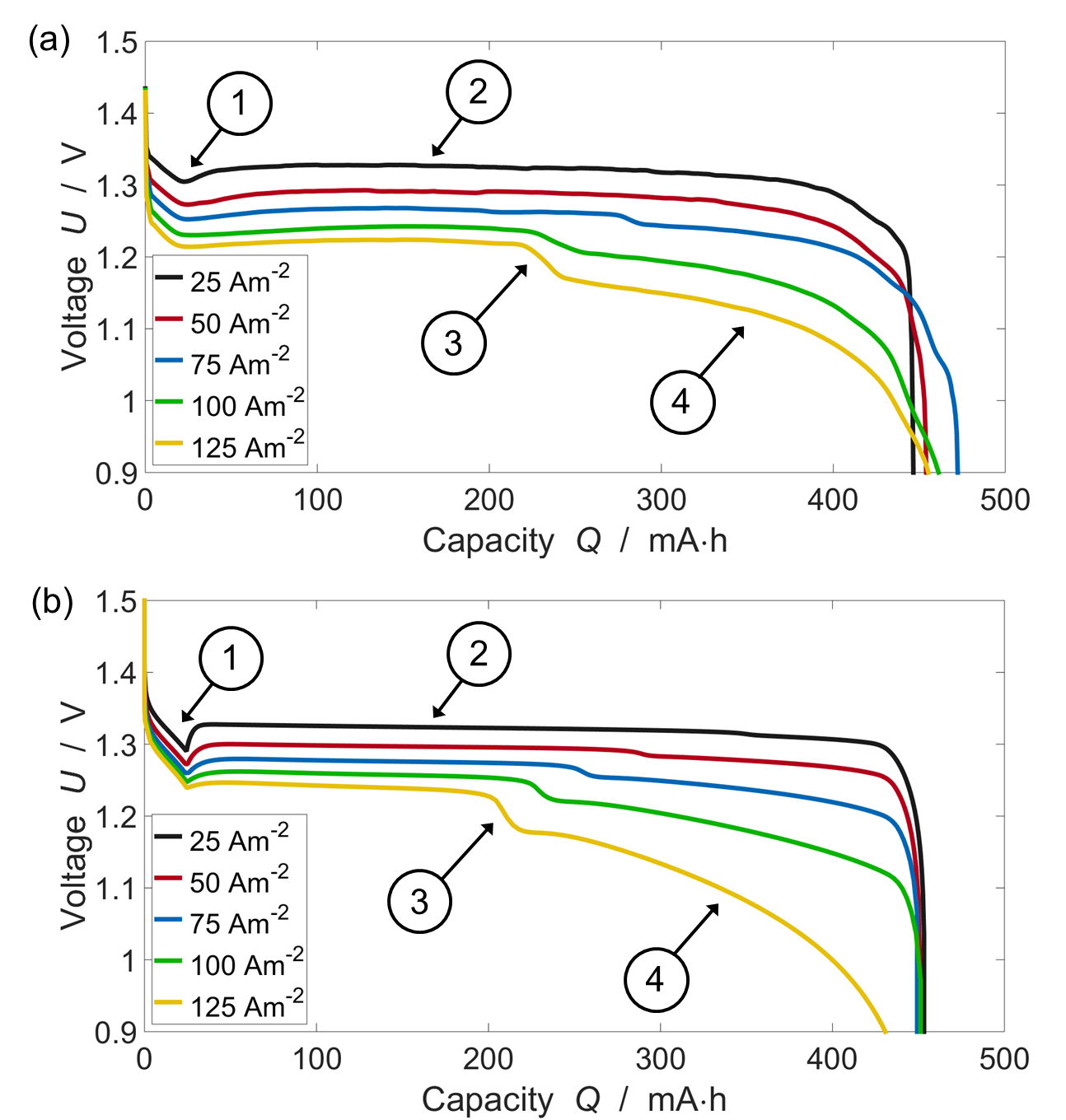}
	  		\caption[Cell voltage during galvanostatic discharge at various current densities (experiment)]{
	(a) Experimentally measured and (b) simulated cell voltage profiles during galvanostatic discharge at various current densities. The four key features in the voltage curves are marked with numbers and discussed in the text 1) initial voltage dip, 2) voltage plateau, 3) voltage step, 4) voltage decay.}
		\label{fig:expshort}
\end{figure}

Discharge profiles at various current densities are plotted in Fig. \ref{fig:expshort}. The voltage generally decreases with increasing discharge current. We observe four characteristic features in each discharge curve:

\begin{enumerate}
\item At the beginning of the discharge process, the voltage drops rapidly until a minimum is reached. Then the voltage recovers slightly. This feature occurs at the same discharged capacity regardless of the current. 

\item After the initial dip, the cell voltage remains nearly constant for more than half of the discharge time. This plateau is wider at smaller discharge currents.

\item At the end of the voltage plateau, the voltage drops rapidly. This voltage step is larger at higher currents and most pronounced at $i=125\text{ Am}^{-2}$. At the smallest current density $i=25\text{ Am}^{-2}$, it is hardly recognizable.

\item After the voltage step, the cell voltage decreases, until it reaches the cut-off voltage.
\end{enumerate}

\subsection{Simulation}

The simulated discharge curves at various currents are depicted in Fig. \ref{fig:expshort}b. We observe the same four characteristics in the simulated discharge profiles as in the experiments (see Sec. \ref{sec:expdischarge}). After a pronounced dip, the voltage remains constant during most of the discharge. The voltage plateau ends with a deep voltage step. Position and magnitude of the step depend on the applied current. Finally, the voltage drops and approaches the cut-off voltage.

\subsubsection{Voltage dip}
\begin{figure}[tb]
\centering
  		\includegraphics[width=\columnwidth]{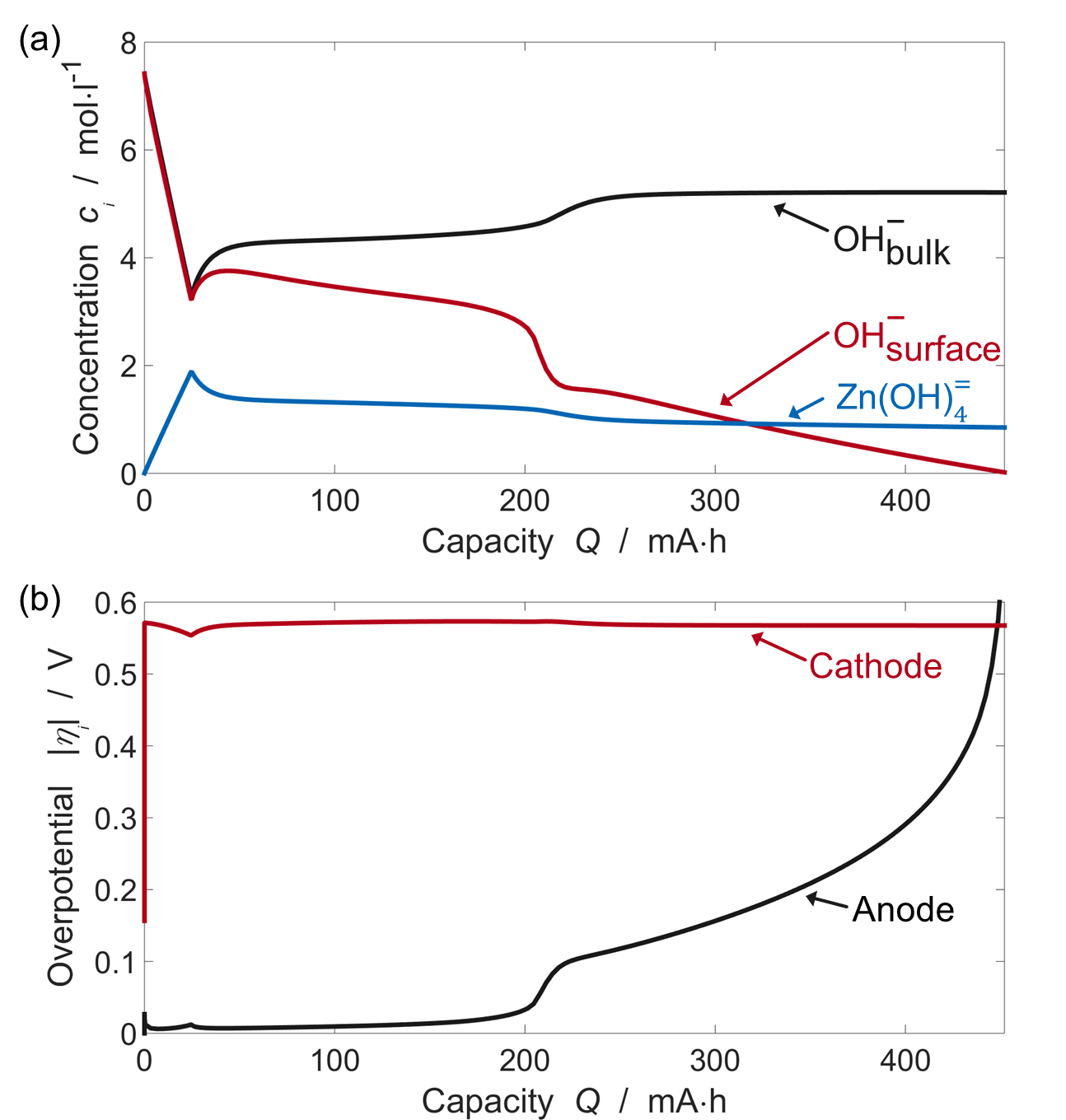}
  		\caption[Mean concentration of hydroxide and zincate]{(a) Mean concentration of hydroxide and zincate in the anode during galvanostatic discharge at 125 Am$^{-2}$. The zincate concentration rises until the critical supersaturation is reached consuming more hydroxide than replaced by oxygen reduction. Then ZnO nucleates and precipitation starts. In the following, the concentration of zincate decreases slowly due to the growing active surface area of the precipitation reaction. Accordingly, the bulk hydroxide concentration increases again. The hydroxide concentration at the Zn surface is dropping at the end of discharge due to the diffusion through the ZnO shell. (b) Overpotentials during galvanostatic discharge at 125 Am$^{-2}$. The high cathodic overpotential remains nearly constant during the discharge process. It is the origin of the initial activation overpotential in the voltage profile (see Fig. \ref{fig:expshort}b) and thereby the plateau voltage. The final growth of the anodic overpotential originates from the loss of hydroxide concentration at the Zn surface  (see Fig. \ref{fig:meanconc}a) and determines the drop in the cell voltage profile.}
		\label{fig:meanconc}
\end{figure}

First, we discuss the initial voltage dip for the discharge current $i=125\text{ Am}^{-2}$. Activation of the slow oxygen reduction leads to a sharp instantaneous voltage drop from the OCV. The voltage dip is a signature for the nucleation of ZnO. We illustrate this based on the mean ion concentrations in the anode (see Fig. \ref{fig:meanconc}a).  

Fig. \ref{fig:meanconc}a depicts the mean concentrations of hydroxide and zincate ions. Initially, the zincate concentration increases linearly, while the hydroxide concentration decreases linearly. Subsequently, the zincate concentration decreases slightly, while the hydroxide concentration increases. Then the hydroxide concentration at the Zn surface decreases, while the bulk hydroxide concentration increases.

During Zn oxidation, hydroxide is consumed and zincate is formed. This explains the initial linear increase in zincate concentration and the decrease in hydroxide concentration. A lower hydroxide and a higher zincate concentration result in a larger overpotential in the anode. When the critical supersaturation is reached, ZnO starts to nucleate and zincate precipitates as ZnO. With increasing area for precipitation, the zincate concentration decreases and the hydroxide concentration increases. This results in an increase in cell voltage. We will explain the final drop in hydroxide concentration at the Zn surface with the diffusion through ZnO below and correlate it to the cell potential.

\subsubsection{Voltage plateau}
After the nucleation of ZnO, the battery discharges in a quasi-stationary regime as expected for conversion reactions. Only the amounts of Zn and ZnO change and affect the surface areas in the anode. This stationary regime is clearly demonstrated by the cathodic and anodic overpotentials during discharge shown in Fig. \ref{fig:meanconc}b. The instant reaction activation overpotentials are very pronounced. The cathodic overpotential dominates and remains constant throughout the full discharge. In the plateau region, the anodic overpotential increases slightly before it rapidly increases after the voltage step. This final increase is due to the diffusion limitation through ZnO (see below).

\subsubsection{Voltage step and decay}
\begin{figure*}
\centering
  		\includegraphics[width=0.9\textwidth]{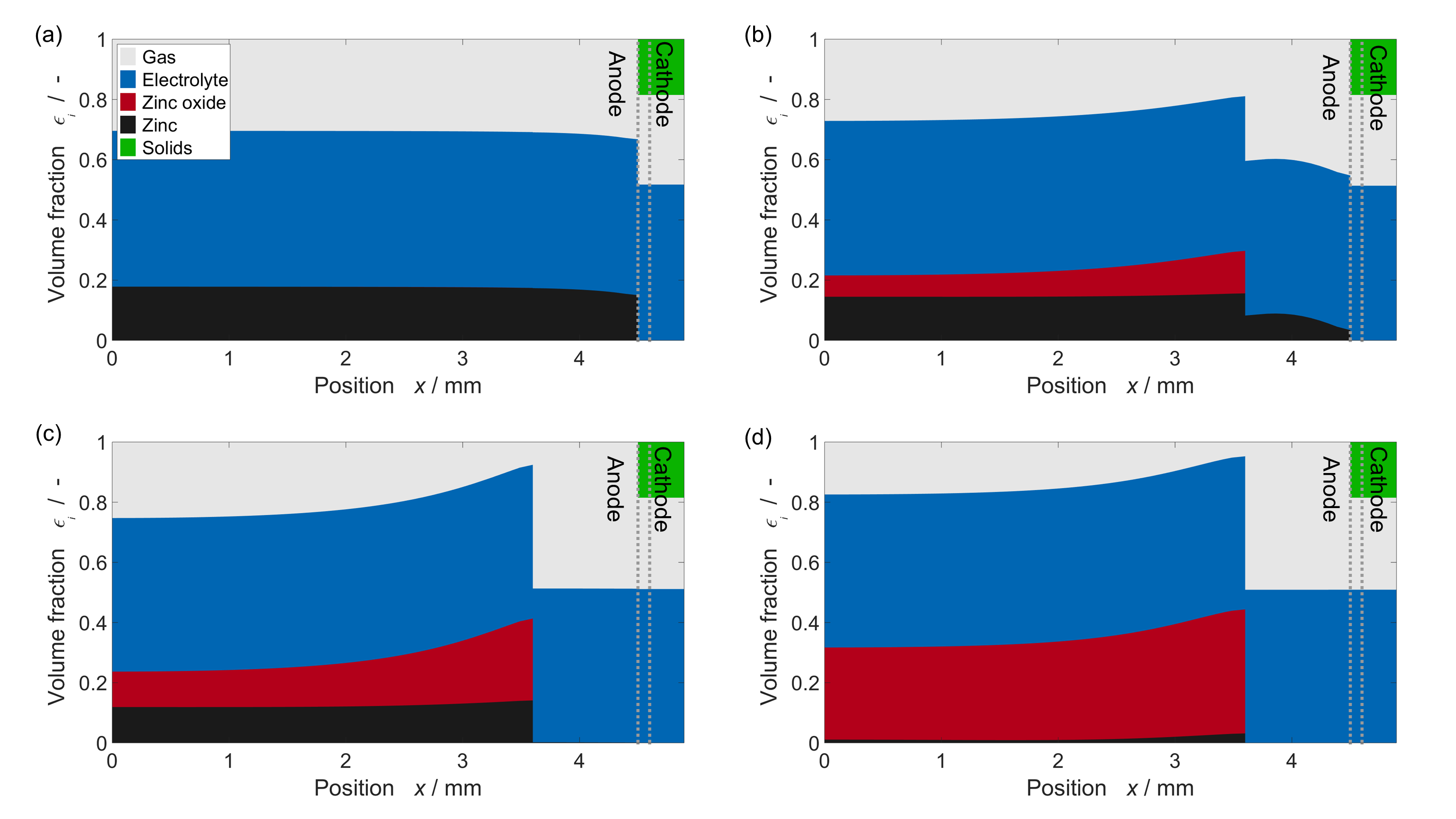}
  		\caption[Volume fractions at characteristic times]{Volume fractions during galvanostatic discharge at 125 Am$^{-2}$ at characteristic times (see Fig. \ref{fig:expshort}): (a) Dip: No ZnO is precipitating. Zn dissolves slightly faster next to the separator. (b) Plateau: ZnO nucleated and precipitating in the part of the anode close to the current collector (see Fig. \ref{fig:charconc}). Zn dissolution is slowed down in the presence of ZnO. (c) Step: Zn is completely dissolved in the part of the anode in which no ZnO is nucleated. (d) Drop: A thick ZnO film slows down the dissolution of the remaining Zn.}
		\label{fig:volfrc}
\end{figure*}

\begin{figure*}
\centering
  		\includegraphics[width=0.9\textwidth]{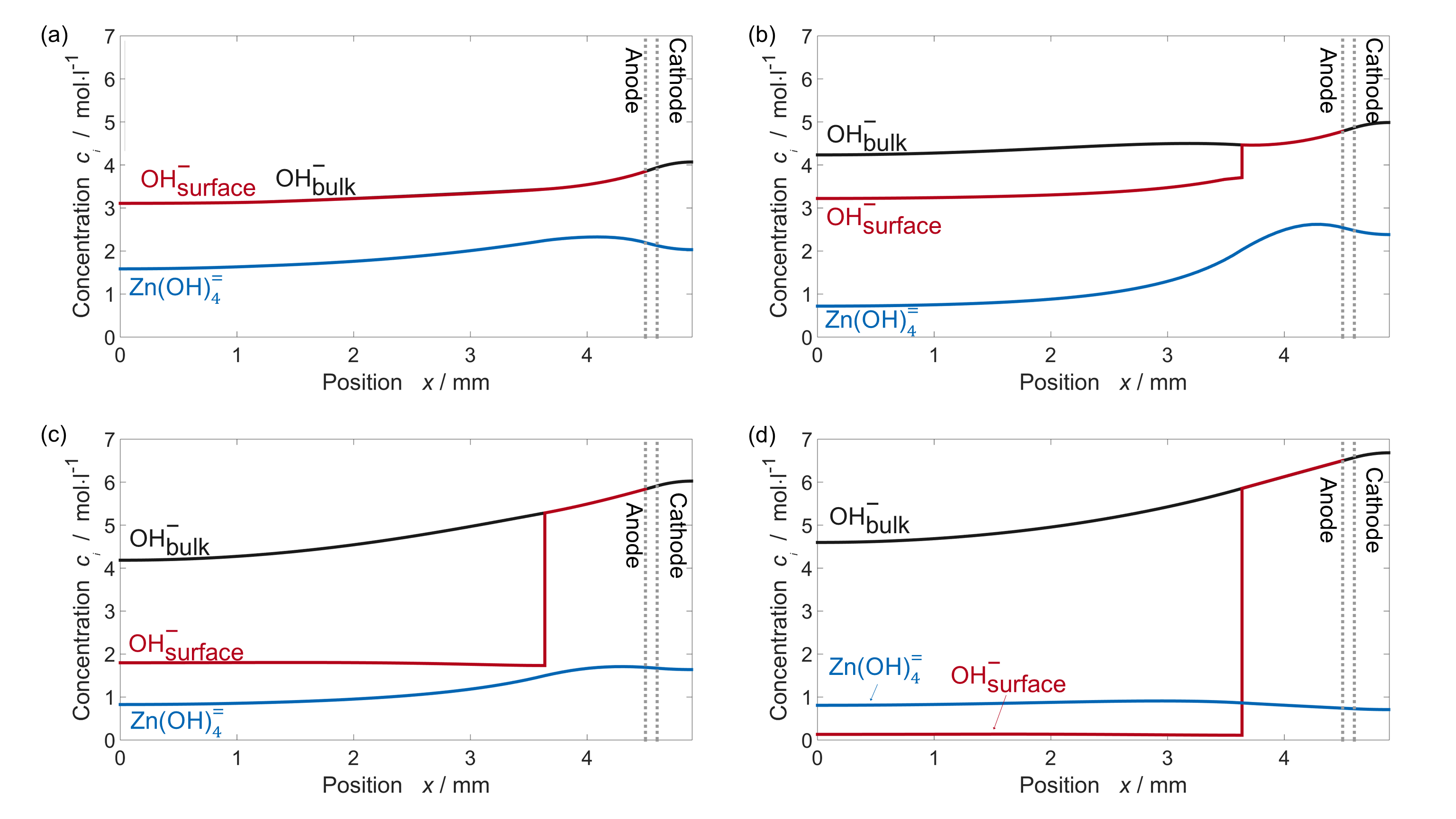}
  		\caption[Concentration profiles at characteristic times]{Various concentration profiles during galvanostatic discharge at 125 Am$^{-2}$ at characteristic times (see Fig. \ref{fig:expshort}): (a) Dip: Zincate concentration is maximum. Critical supersaturation is reached next to the current collector, where zincate solubility is low due to low potassium concentration. (b) Plateau, and (c) Step: Hydroxide bulk concentration and zincate concentration remain nearly constant. Hydroxide surface concentration decreases in parts of the anode due to growing ZnO film. (d) Drop: Hydroxide concentration at the Zn surface is small and limits Zn dissolution.}
		\label{fig:charconc}
\end{figure*}
At the end of the plateau region, a step in cell voltage occurs. Such behavior is typically interpreted with a change in reaction mechanism. The insights from our simulation show, however, that the voltage step originates from the inhomogeneous nucleation of ZnO. 

To explain this mechanism, we plot the volume fractions of all phases in the button cell in Fig. \ref{fig:volfrc}. We observe that ZnO does not nucleate next to the separator, but precipitates next to the current collector. There a ZnO film passivates the Zn and limits its dissolution. Thus, during the voltage plateau the uncovered Zn close to the separator is preferentially dissolved. When the voltage step is reached, this uncovered Zn is completely gone (see Fig. \ref{fig:volfrc}c). The remaining Zn is already covered with a thick ZnO film. After the voltage drop, the Zn beneath this thick film takes over and its oxidation becomes responsible for the cell current. These observations explain the voltage step. It results from the sudden change from the oxidation of uncovered to covered Zn that requires a jump in the driving force for oxidation.

We highlight the importance of electrolyte management based on Fig. \ref{fig:volfrc}. Initially, a huge void space filled with gas is present throughout the anode. This is a model representation of the gas space on top of the anode in the real button cell. During discharge the solid volume fraction increases as Zn is converted into ZnO. The void space ensures that the electrolyte is not leaking out of the gas diffusion electrode. This void space just stays open at the end-of-discharge demonstrating that this VARTA button cell is well optimized. {In conclusion, we find that the electrolytic parameters remain stable during battery discharge and charge. Furthermore, a well optimized gas diffusion electrode guarantees stable oxygen concentrations throughout the cell at these relatively low current densities, as shown previously \cite{Horstmann2013}.}

Next, we want to understand the origin of this inhomogeneous precipitation from the concentration profiles depicted in Fig. \ref{fig:charconc}. It shows hydroxide concentration at the electrode surface and in the bulk together with zincate concentration. Note that the potassium concentration is the sum over these anionic concentrations. We observe that the hydroxide concentration is increasing towards the cathode whereas the zincate concentration is maximal in the anode close to the separator. At the voltage dip, the rising zincate concentration surpasses the supercritical concentration next to the current collector where potassium concentration and zincate solubility are lowest (see Supplementary Materials A1.2). Next to the separator, potassium concentration and solubility are higher and do not allow ZnO nucleation. 

During further discharge, the concentration of hydroxide at the Zn surface is decreasing in the presence of the ZnO film (see \ref{fig:charconc}d and Fig. \ref{fig:meanconc}a). This causes the final cell voltage loss and increase in anode overpotential at 125 Am$^{-2}$. The ZnO film is growing during discharge and acts as a diffusion barrier for hydroxide (see Eq. \ref{portrans}).  

\subsubsection{Discharge currents}
In this section, we compare the simulated discharge curves for various current densities (see Fig. \ref{fig:expshort}b). It is clear that reaction rates, transport rates, and thus the overpotential depend on the discharge current. Higher currents generally result in higher overpotentials. In Fig. \ref{fig:expshort}b, on closer examination it is found that only the discharge curve at 125 Am$^{-2}$ shows the typical shape for diffusion limitations. We find in our simulations that the discharge at lower currents is limited by the total Zn amount in the anode. 

The voltage step is an important finding of this paper. The discharged capacity at which this step occurs decreases with increasing discharge current. This can be explained by comparing the profiles of the specific active surface areas for ZnO precipitation in Fig. \ref{fig:meanconc2}a. The active surface areas increase from current collector to separator in the anode. Next to the separator, no surface area is available. This region increases with increasing discharge current.

The increase in surface area from the current collector to the separator shows that ZnO growth is preferred close to the separator and leads to larger ZnO shells. The absence of surface area next to the separator is explained above via inhomogeneous nucleation. A faster discharge leads to a larger overshoot of the zincate concentration above the supercritical limit, which increases the nucleated surface. Thus, the region without ZnO is smaller and the voltage step is observed at a lower discharged capacity.

\subsubsection{Battery cycling}
\begin{figure}[h]
\centering
  		\includegraphics[width=\columnwidth]{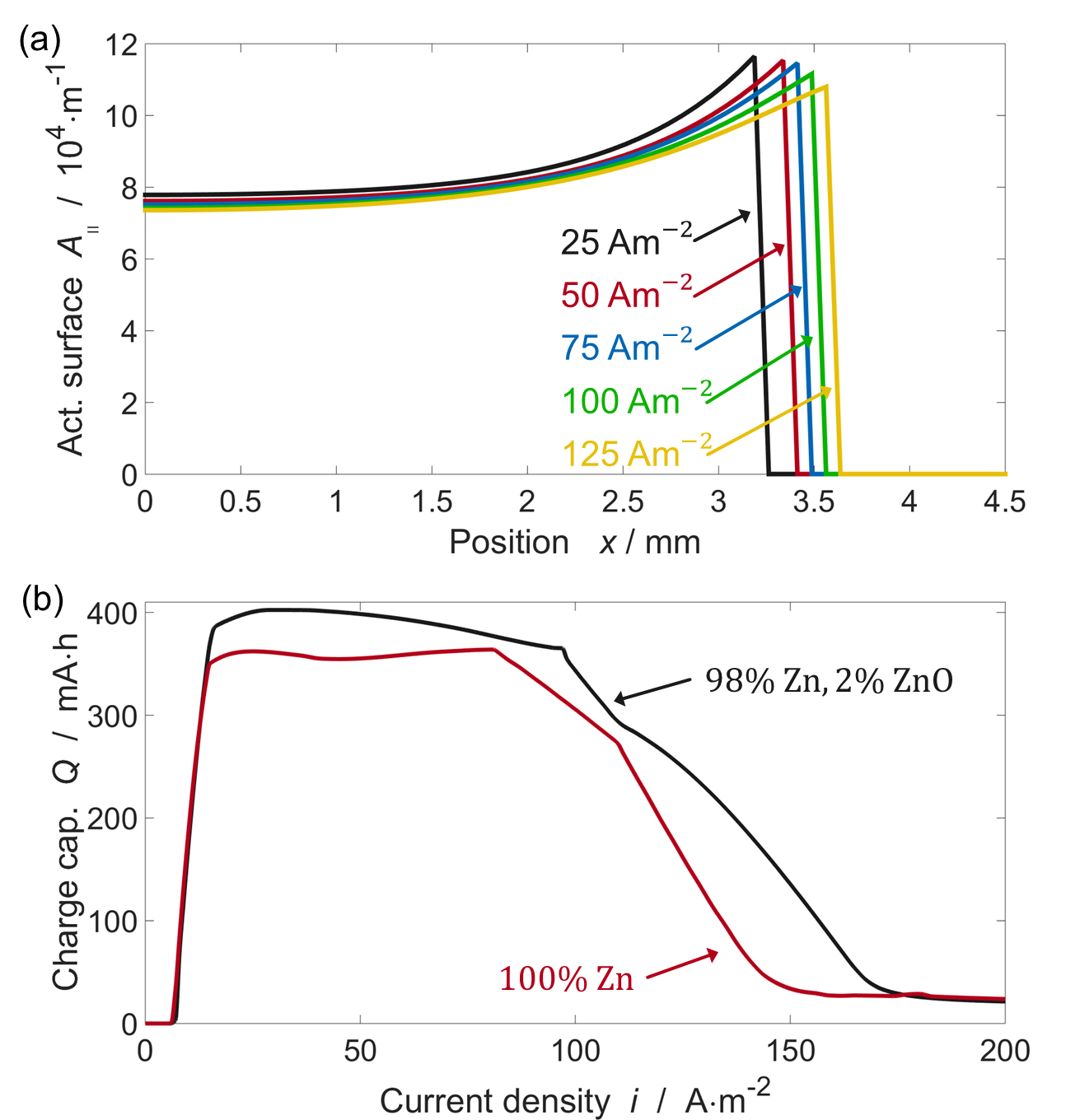}
  		\caption[Active surface of ZnO precipitation]{(a) Active surface for ZnO precipitation during galvanostatic discharge at various current densities. For higher current densities the region where no ZnO precipitates is smaller. Therefore, the capacity at which all Zn is dissolved in this region is smaller for higher currents and the voltage step occurs at lower capacities. {(b) Rechargeable capacity after discharge to $Q=407\text{ mAh}$ (90\% capacity) or $U=1.1\text{ V}$. By mixing ZnO into the Zn anode, the rechargeable capacity increases.}
		}
		\label{fig:meanconc2}
\end{figure}
{The implications of our analysis for the development of rechargeable zinc-air batteries are demonstrated by simulating one recharge after a full discharge to the fixed capacity $Q=407\text{ mAh}$ or the voltage cut-off $U=1.1\text{ V}$ (see Supplementary Materials D). We consider two scenarios: First, we prepare a pure Zn anode; second, we admix 2 volume percent ZnO. In the latter case the voltage dip during discharge disappears. This is because the admixture of ZnO makes its nucleation needless and guarantees more homogeneous deposition of ZnO. As a consequence, Zn dissolves more homogeneously.

We compare the corresponding rechargeable capacities in Fig. \ref{fig:meanconc2}b. Let us explain the different regimes for the example of a pure Zn anode. At very low current densities $i\lesssim 15\text{ Am}^{-2}$, $\text{CO}_2$ absorption limits the capacity. At intermediate current densities $15\text{ Am}^{-2}\lesssim i\lesssim 80\text{ Am}^{-2}$, the rechargeable capacity remains almost constant because it is limited by the amount of accessible ZnO. At $80\text{ Am}^{-2}\lesssim i\lesssim 110\text{ Am}^{-2}$, dissolution of ZnO is capacity limiting even though ZnO remains available in the anode. Above $i\gtrsim 110\text{ Am}^{-2}$, discharge capacity is limiting. We observe that the admixture of ZnO increases the rechargeable capacity because ZnO and Zn remain evenly distributed. This effect is most pronounced at relatively low currents where the surface area for ZnO dissolution is capacity limiting. To conclude, admixture of ZnO leads to more homogeneous precipitation/dissolution and increases the rechargeable capacity of zinc-air batteries.}

\subsection{Validation and Discussion}
Finally, we compare measured (see Fig. \ref{fig:expshort}a) and simulated discharge (see Fig. \ref{fig:expshort}b). Note that this paper highlights the qualitative agreement of our theory-based continuum modeling with electrochemical measurements and gives novel insight based on simulations. Excellent quantitative agreement, which is not the aim of this paper, could be gained by adding more parameters and performing extensive parameter adjustments. Nevertheless, we discuss potential model refinements in the following. The four characteristic features, i.e., voltage dip, voltage plateau, voltage step, and voltage drop, are found in both, theory and experiment.

Around the voltage dip, the cell potentials do not exactly agree. Activity coefficients are not included in our modeling, but would affect the cell potential when the zincate concentrations is supercritical at the voltage dip. The increase of the voltage after the dip is very sharp in our simulations. This might be a result of our mean-field description of the nucleation process assuming that nucleation happens in one burst. In reality, small islands might nucleate, grow, and merge to form closed ZnO shells. In this case of agglomeration, the specific surface area for ZnO precipitation would approach full coverage more continuously. This effect would lead to a slower increase in cell voltage.

Magnitude and position of the voltage step agrees very well between model and measurement. At small current densities the voltage step is barely visible in the measured discharge curves, whereas it can still be observed in the simulation. This is a consequence of the lower noise level in the simulations. This excellent agreement supports our interpretation that the voltage dip signals inhomogeneous nucleation. In secondary zinc-air cells it would be advantageous to precipitate ZnO and dissolve Zn homogeneously. This can be achieved by preparing the anode as a mix of ZnO and Zn. The added ZnO will act as nucleation seed for precipitation. This method will reduce the initial discharge capacity, but improve the cycle life of the Zn anode.

The final diffusion limited voltage drop represents a significant shortcoming of our model. In simulations, the diffusion limited regime is only found at 125 Am$^{-2}$, whereas in experiments, it seems to occur at smaller currents, too. It is interesting that the measurements show a non-monotonous dependence of total discharge capacity on cell current. Furthermore, this behavior at the end of discharge is not exactly reproducible in our experiments. This indicates that the end of discharge is influenced by degradation. Degradation strongly depends on lab conditions, e.g., moisture, temperature. Examples of degradation mechanisms are hydrogen evolution, corrosion of the current collectors, and formation of type II Zn. Type II Zn is not contained in our model but would create an additional diffusion barrier at low enough voltages $U\lesssim 1.1 \text{ V}$ at the end of discharge \cite{Horn2003}. Also, taking into account a Zn radius distribution would result in a smoother decay of Zn surface area at the end of discharge and a smoother decay in cell voltage as observed in our measurements \cite{Horstmann2013,Rinaldi2016}. We capture diffusion limitations at the end of discharge by the simplified rate Eqs. \ref{portrans} and \ref{znopercreact} for Zn dissolution and ZnO precipitation and propose microscopic modeling of electrolyte transport around individual Zn particles.

\section{Lifetime Analysis}
\label{sec:lifetime}
We study the lifetime of the zinc-air button cell (see Sec. \ref{sec:setup} for measurement sequence) in this section. To this aim, we perform experiments and simulations giving insights into the underlying degradation process. Our measurement procedure is described in detail in Sec. \ref{sec:expdischarge}. First, we  prepare the cell through a continuous discharge and let ZnO nucleate. Every following day we measure the steady-state voltage for the current densities 100 Am$^{-2}$ or 50 Am$^{-2}$. When the cut-off voltage of 0.9 V is reached, we finish our measurement. The measured voltage as a function of time is shown in the Supplemental Materials in in Fig. B.1.

\begin{figure}[tb]
\centering
\includegraphics[width=\columnwidth]{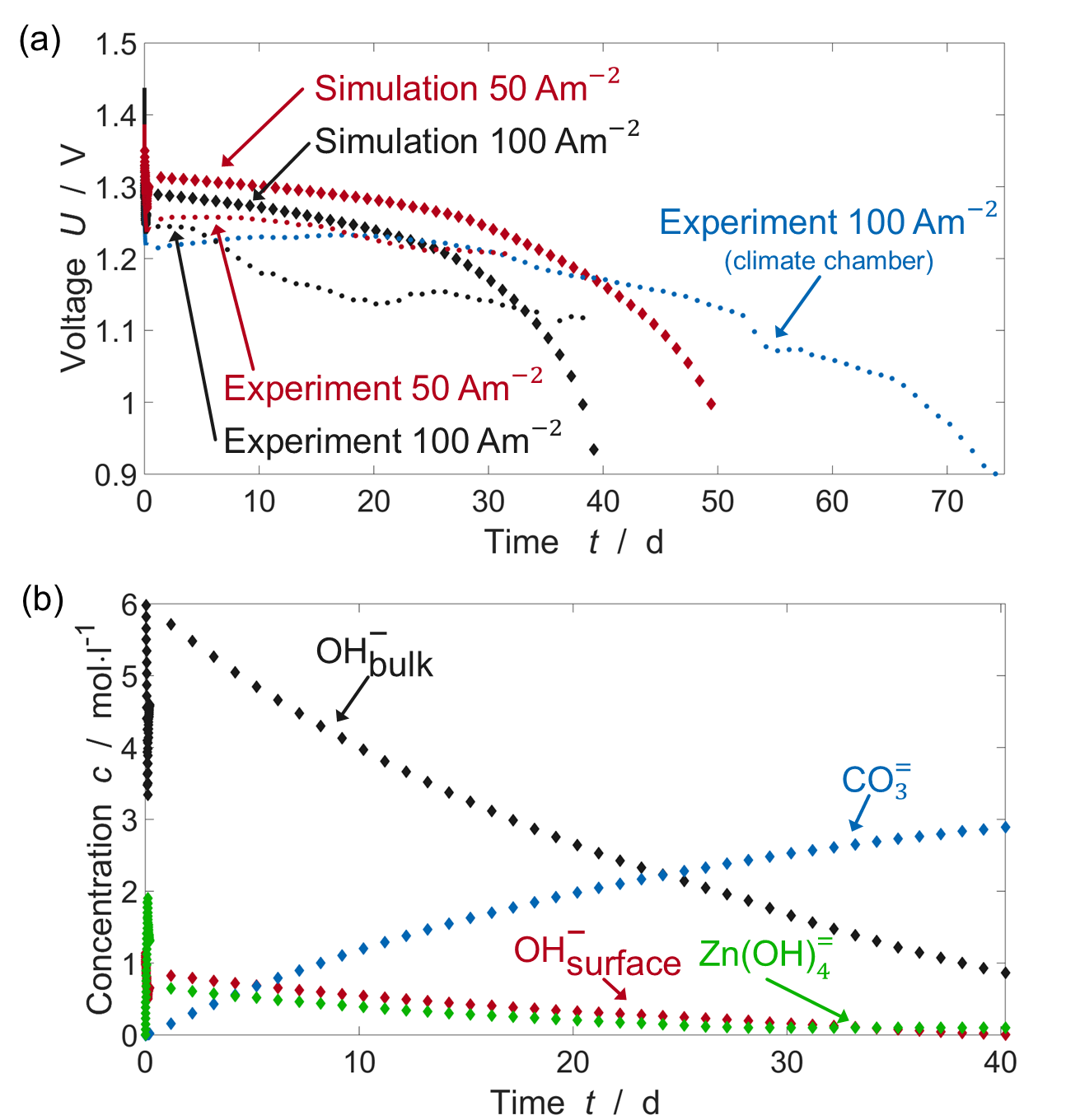}
\caption[Voltage profiles of lifetime experiments]{
Lifetime analysis at 50 Am$^{-2}$ (red) and 100 Am$^{-2}$ (black) (a) Cell voltage profile in lifetime experiment (diamonds) and simulation (dots). One experiment (blue) is carried out in the climate chamber at 20$^\circ$C, while the other experiments are performed at room temperature and at lab atmosphere. 
The initial voltage drop is caused by the 5h galvanostatic discharge. Thereafter, the cell voltage decreases linearly over time. The decrease is independent from the current density. At the end of battery lifetime, faster voltage decay occurs. (b) Mean concentrations of various ions during lifetime experiment at 100 Am$^{-2}$. The hydroxide concentration $c_{\text{OH}^-}$ is decreasing due to carbonate $c_{\text{CO}_3^{=}}$ absorption. This reduces the zincate solubility/concentration $c_{\text{Zn(OH)}_4^{=}}$. The reduction of hydroxide concentration slows down further Zn dissolution and limits the cell lifetime.}
\label{fig:longexp}
\end{figure}

Various aging profiles are plotted in Fig. \ref{fig:longexp}a. During the daily measurements, the voltages increase slightly for around $10$ days before starting to decrease slowly. Accelerated voltage decay and cell failure occur after one to two months. The voltage profiles are not reproducible and contain a significant amount of noise. In contrast, during short galvanostatic discharge, the voltages are well reproducible. We attribute this noise to the fluctuating environmental conditions in our lab, e.g., temperature, pressure, and air composition. A reference measurement in a climate chamber yields smoother results, even though the air composition is not controlled.

Lifetime simulations are depicted in Fig. \ref{fig:longexp}a. The voltage is decreasing within a day in our simulations before decreasing logarithmically. The lifetime is limited to 40 days. We study the origin of cell failure by plotting mean ion concentrations in Fig. \ref{fig:longexp}b. The absorption of carbon dioxide and formation of carbonate leads to an almost linear increase in carbonate concentration (see Reaction \ref{co2abs}). Because this carbonate formation consumes hydroxide, the hydroxide concentration in the electrolyte is reduced significantly. A low hydroxide concentration results in low zincate solubility. The battery cell finally fails due to this decrease in hydroxide concentration and zincate solubility which slow down the further dissolution of Zn. Note that this mechanism does not involve the precipitation of solid carbonates as shown for alkaline electrolytes before \cite{Ko1983}. Instead, the reduction in pH is the major consequence of carbon dioxide absorption and the cause for cell failure.

{A strategy for mitigating carbon dioxide absorption is illustrated in the Supplementary Materials C. We simulate how a decrease in carbon dioxide content in the feed gas extends the lifetime. Drillet et al. analyze the use of carbon dioxide filters to this aim \cite{Drillet2001}.}

We now compare the simulated voltage profile with the measured one (see Fig. \ref{fig:longexp}a). Generally, the simulated voltages are $50\text{ mV}$ too large. This deviation is less distinct during galvanostatic discharge (see Sec. \ref{sec:discharge}). It stems from the relaxation in zincate and hydroxide concentration during battery storage. Agreement would be improved by substituting the global electrochemical kinetics (see Eqs. \ref{butvola},\ref{redrate}) with adjusted expressions \cite{Sunu1980, Bockris1972, Cao2003} that weaken the dependence of reaction kinetics on ionic concentrations. The good agreement in the voltage slope and the lifetime shows that our simulations qualitatively capture the lifetime limitation. This level of agreement is reached by using a relatively small surface area for carbonate formation, which is two orders of magnitude smaller than expected. Previous studies show a shorter battery lifetime \cite{Drillet2001,Schroder2014} of around 10 days. Therefore, we hypothesize that the measured VARTA button cell is optimized to reduce carbon dioxide absorption.

\section{Conclusion}
\label{sec:conclusion}
{Zinc-air batteries were proposed as promising candidates for stationary energy storage due to the use of abundant materials. In this article, we model the discharge of a commercial zinc-air button cell and validate it with experiments. Our simulations describe electrolyte convection and take into account nucleation and growth of the discharge product. 
We find that the primary zinc-air battery exhibits inhomogeneous deposition and dissolution of ZnO and Zn. Adding ZnO to the zinc anode is shown to improve the rechargeable capacity even though it reduces the initial discharge capacity. Additionally, we show that battery lifetime is limited by carbon dioxide absorption into the aqueous alkaline electrolyte. This effect can be mitigated by using carbon dioxide filters or employing neutral electrolytes.}

\section{Acknowledgement}
The authors thank Martin Krebs (VARTA Microbattery) and Simon Clark for fruitful discussions. We acknowledge financial support by the EU commission through the project ZAS! Zinc-Air Secondary (Grant Agreement 646186). Further support was provided, by the bwHPC initiative and the bwHPCC5 project through associated compute services of the JUSTUS HPC facility at the University of Ulm.

\bibliographystyle{elsarticle-num}
\bibliography{library}

\end{document}